\documentclass[final,5p,twocolumn]{elsarticle}

\pdfoutput=1
\usepackage[utf8]{inputenc}
\usepackage{epsf}
\usepackage{epsfig}
\usepackage{graphics}
\usepackage{graphicx}
\usepackage{amssymb}
\usepackage{amsmath}
\usepackage{multirow}
\usepackage{color}
\usepackage{ulem}

\newcommand{\nc}{\newcommand}
\nc{\postscript}[2] 
{\setlength{\epsfxsize}{#2\hsize}\centerline{\epsfbox{#1}}}
\nc{\non}{\nonumber}
\nc{\hc}{\hbox {h.c.}} \nc{\re}{\hbox {Re}} 
\nc{\mev}{\hbox {MeV}} \nc{\gev}{\;\hbox {GeV}} \nc{\tev}{\;\hbox {TeV}}
\def\lsim{\mathrel{\raise.3ex\hbox{$<$\kern-.75em\lower1ex\hbox{$\sim$}}}}
\def\gsim{\mathrel{\raise.3ex\hbox{$>$\kern-.75em\lower1ex\hbox{$\sim$}}}}

\nc{\etal}{{\it et al.}}
\nc{\Lsp}{\;\;\;\;\;\;\;\;\;\;}  \nc{\LLLsp}{\lspace \lspace}
\nc{\lsp}{\;\;\;\;\;\;}
\nc{\spac}{\;\;\;}
\nc{\noi}{\noindent}
\nc{\beq}{\begin{equation}}   \nc{\eeq}{\end{equation}}
\nc{\bea}{\begin{eqnarray}}   \nc{\eea}{\end{eqnarray}}
\nc{\baa}{\begin{array}}      \nc{\eaa}{\end{array}}
\nc{\bit}{\begin{itemize}}    \nc{\eit}{\end{itemize}}
\nc{\ben}{\begin{enumerate}}  \nc{\een}{\end{enumerate}}
\nc{\bce}{\begin{center}}     \nc{\ece}{\end{center}}
\nc{\red}{\textcolor{red}}

\def\sq2{\sqrt{2}}

\def\ph{\varphi}

\def\m4{m^4(\ph)}
\def\mn2{M_n^2}

\def\v5{V^{(5)}}

\journal{Physics Letters B}

\begin{document}

\vspace{.5cm}

\title{ 
Additional Higgs bosons: Supersymmetry or warped extra dimensions?}
\author[1]{Mariana Frank}
\ead{mariana.frank@concordia.ca}
\author[2]{Nima Pourtolami}
\ead{nima.pourtolami@gmail.com}
\author[3]{Manuel Toharia}
\ead{mtoharia@dawsoncollege.qc.ca}

\affiliation[1]{organization={Department of Physics,
  Concordia University}, addressline={7141 Sherbrooke St. West},
  city={Montreal}, postcode={Quebec H4B 1R6}, country={Canada}}

\affiliation[2]{organization={National Bank of Canada}, addressline={1155 Metcalf St.},
  city={Montreal}, postcode={Quebec H3B 4S9}, country={Canada}}

\affiliation[3]{organization={Department of Physics, Dawson College},addressline={3040 Sherbrooke
  St W}, city={Montreal}, postcode={Qc, H3Z 1A4}, country={Canada}}

\date{\today}


\begin{abstract}

We investigate and compare additional CP-even, CP-odd and charged
scalar states appearing in two popular Beyond the 
Standard Model scenarios. We focus on the simplest possible Higgs sector
within warped extra-dimensions and supersymmetry, with the aim to differentiate between them. 
In each case, we analyze  the couplings of the new
Higgs states, 
looking for distinguishing signatures.
We show that the couplings of the Standard Model gauge bosons to the first Kaluza-Klein Higgs
states of the extra-dimensional setup (CP-even, CP-odd and charged) 
are very similar to those of the heavy Higgs states of the
MSSM in the decoupling region. We also find that the Yukawa
couplings in the extra-dimensional scenario can mimic the different types of Yukawa couplings
of general Two-Higgs Doublet Models, in particular the so-called
Type-II couplings, which are similar to those in the MSSM.

\end{abstract}

\maketitle

\section{ Introduction}
The Large Hadron Collider (LHC) was specifically designed to detect the
elusive Higgs boson of the Standard Model (SM). After the initial
elation following its discovery \cite{ATLAS:2012yve, CMS:2012qbp}, given the
shortcomings of the SM, theorists and experimentalists alike
are searching for Beyond the SM (BSM). Given LHC's high sensitivity to Higgs-like particles, whether the new physics will indicate new
particles or new interactions, it is likely that 
additional Higgs bosons will be among the first to be detected\cite{Lykken:2010mc}. With this assumption, the main question would be  to establish the
underlying structure responsible for these states.  
Our aim here is to discuss and compare the predictions for the Higgs sector of 
two popular BSM scenarios: supersymmetry and warped extra-dimensions.

In the case of supersymmetry, we assume that supersymmetric
partners are heavy enough to suppress their direct production at the LHC.
In that situation the scenario becomes effectively a Type II Two Higgs Doublet Model (2HDM) \cite{Branco:2011iw}, with some additional constraints coming from the supersymmetric structure. 
An implementation of this framework is for example the hMSSM, where the mass of the SM-like 
  Higgs boson is fixed to be the measured value from the LHC,
  $M_h=125$ GeV. This is accomplished by restricting the amount of
  radiative corrections to its mass, yielding as a result, a very
  heavy supersymmetric particle spectrum
  \cite{Arcadi:2022hve,Djouadi:2013uqa,Djouadi:2013vqa,Djouadi:2015jea,Bagnaschi:2021jaj}. 

Warped extra-dimensional models \cite{Randall:1999ee} can accommodate a minimal
implementation of the Higgs mechanism within the bulk of the
extra-dimension through a single 5-dimensional Higgs
doublet, invariant under the SM gauge group. With this minimal gauge structure, 
the Higgs sector of the warped scenario contains, in addition to a SM-like Higgs,
complete towers of CP-even, CP-odd and charged Kaluza-Klein (KK) Higgs
states. These scenarios, however, are typically very constrained by electroweak precision data and
flavour-changing neutral currents (FCNCs) so that the masses of the
lightest KK states must be pushed into $\cal{O}$(10) TeV scales or higher
\cite{Burdman:2002gr, Agashe:2003zs}. By appropriately modifying the warped background
metric \cite{Cabrer:2011fb}, and/or adding brane kinetic terms for the
gauge fields, \cite{Carena:2002dz} one can evade flavour and electroweak precision constraints for relatively
light KK states, without having to extend the gauge or the Higgs
sectors.  For the purposes of this work, we consider the special
situation in which the lightest  KK states  are the
KK Higgs states of $\cal{O}$(1) TeV \cite{Frank:2016vtv}, so that they could be produced
and observed at the LHC. 

As each of these two classes of models have promising theoretical features,  in the event that additional scalars are discovered, it would be essential
to know if the lowest KK Higgs states of warped extra dimensions (CP-even, CP-odd and charged) can be distinguished from their counterparts in
MSSM.   
%
%
We start by a brief description of each model.
\section{Minimal Higgs sector in a 5D warped model}
We consider a scenario with one extra space dimension and assume a
(properly stabilized) static spacetime background as 
\bea
ds^2 = e^{-2\sigma(y)}\eta_{\mu\nu} dx^\mu dx^\nu - dy^2\, ,
\label{RS}
\eea
where $\sigma(y)$ is a warp factor
responsible for the exponential suppression of the mass scales from the UV brane, down to the IR brane, located at the two boundaries of the extra coordinate, $y=0$ and $y=y_{1}$, respectively 
\cite{Randall:1999ee,Randall:1999vf}.

The matter content of the model corresponds to a minimal 5D extension of the
SM, with the same gauge groups $SU(3)\times SU(2)\times U(1)$, and with all
fields propagating in the bulk
\cite{Davoudiasl:1999tf,Grossman:1999ra,Pomarol:1999ad}, such that the localization of 
fermions can  resolve the flavor puzzle of the SM \cite{Agashe:2004cp}.
The electroweak symmetry breaking (EWSB) is induced by a single 5D bulk Higgs
doublet appearing in the electroweak Lagrangian density as
\bea
&&\hspace{-1cm}{\cal L}=\sqrt{g}\ \left(-\frac{1}{4} F^2_{MN} +
|D^M H|^2 - V(H) \right)  \, , \non\\
&&\hspace{-1.5cm} + \sum_{i=1}^{2}\sqrt{g}\ \delta(y-y_i)
\left(-\frac{r_i}{4 k} F^2_{MN} +\frac{d_i}{k}  |D^M H|^2 - \lambda_i(H)\right)\!,  
\label{eq:5Daction}
\eea
where the capital indices $M, \, N, ...$ are used to denote the $5$ spacetime
directions, while the 
Greek indices $\mu, \, \nu, ...$ are exclusively used for the 4D directions.

The 5D action in Eq. (\ref{eq:5Daction})
includes possible brane-localized kinetic terms associated with the
gauge fields and the 5D Higgs, which are proportional to
$\delta(y-y_i)$.
The strength of these terms are characterized by the free parameters, $r_i$ and $d_i$ (in units of $k \sim M_{Pl}$).
These terms alter the predicted spectrum of the Higgs 
KK modes, affecting the masses of the lightest CP-odd and charged KK
Higgs, and that of the second lightest CP-even KK Higgs \cite{Frank:2016vtv}\footnote{The mass of the lightest mode, being the SM
Higgs, can still be fixed to its observed value by adjusting the coefficients of the quartic term of the brane Higgs potential.}.

The 5D Higgs doublet can be expanded around a nontrivial VEV profile
$v_5(y)$ in a similar way as in the SM
\bea
\label{Hexpansion}
H=\frac{1}{\sqrt{2}} e^{i g_5\Pi  }\left(\begin{matrix} 0\\ v(y) +h(x,y) \end{matrix}
\right) \, ,
\eea
with the covariant derivative being $D_M= \partial_M + i g_5 \mathbb{A}_M$, where $\mathbb{A}_M$ is the gauge field
\bea
\label{Aexpansion}
\mathbb{A}_M=\left(
\begin{matrix}
  s_W \mathbb{A}_M^{em}+\frac{c_W^2-s_W^2}{2c_W}Z_M & \frac{1}{\sqrt{2}}W_M^+\\
 \frac{1}{\sqrt{2}}W_M^-& -\frac{1}{2c_W} Z_M  
 \end{matrix}
\right) \,.
\eea
The CP-odd, $A$, and charged Higgs, $H^{\pm}$, degrees of freedom are contained in
\bea
\label{Piexpansion}
\Pi=\left(
\begin{matrix}
\frac{c_W^2-s_W^2}{2c_W} A & \frac{1}{\sqrt{2}} H^+\\
 \frac{1}{\sqrt{2}}H^-& -\frac{1}{2c_W} A   
\end{matrix}
\right) \, ,
\eea
($s(c)_W \equiv \sin(\cos) \theta_W$), with a weak angle defined like in the SM, {\it i.e.} $s_W/c_W=g'_5/g_5$,
where $g_5$ and $g_5'$ are the 5D coupling constants of $SU(2)_L$ and $U(1)_Y$.

The extraction of degrees of freedom in this context can be found in
\cite{Falkowski:2008fz,Cabrer:2011fb,Archer:2012qa}, as well as in
\cite{Frank:2016vtv}, where brane kinetic terms are considered. 
In particular, there will be a KK tower of CP-even Higgs bosons, $H_n$, a
KK tower of CP-odd Higgs bosons, $A_n$, and a tower of charged Higgs bosons,
$H^\pm_n$, where $n=0,1,2,..$ denotes the KK mode level, with higher modes
associated to heavier 4D effective masses\footnote{Note that the
  CP-odd physical Higgs KK modes $A_n(x)$ and the 
charged physical Higgs KK modes $H^\pm_n(x)$ will be extracted from a
mixing between the fifth component of the electroweak gauge bosons $Z_5(x,y)$
and $W^\pm_5(x,y)$ and the bulk Higgs bosons $\Pi(x,y)$ and
$H^\pm(x,y)$ (we use $x$  to denote the 4D coordinates and $y$ to
denote the coordinate along the extra dimension).}.

The lightest CP-even KK Higgs boson $H_0$ (which we will denote as $h$) is identified with the 125 GeV boson
 discovered by CERN and its mass scale is fixed by the nontrivial vacuum expectation value (VEV)
 background $v(y)$, which fixes the electroweak scale. The next
 potentially accessible KK Higgs bosons at the LHC would be the second
 CP-even KK Higgs boson $H_1$ (denoted as $H$, in what follows),  the first CP-odd
 KK Higgs $A_0$ (henceforth denoted as  $A$)  and the first  charged KK
 Higgs $H^{\pm}_0$ (henceforth denoted as $H^\pm$). Their masses will be of the order of the warped
 down Planck scale $M_{KK}=e^{- \sigma(y_1)} M_{Pl} \sim {\cal O}(1$ TeV) and their precise
 value will depend on the boundary conditions set by the brane kinetic
 terms. One can show that, up to corrections proportional to the weak
 scale, the masses of these three different KK Higgs bosons will be the
 same, irrespective of the value of the Higgs brane kinetic coefficient
 $d_1$ and the nature of the background metric, {\it i.e.}
 \bea
 M_H^2&=& b_{d_1}\ M_{Pl}^2 e^{-2 \sigma(y_1)}\ \  \sim\ \  {\cal O}({\rm TeV}) \, ,\\
 M_A^2 &=& M_H^2\ \left(1 + {\cal O}(M_Z^2/M_{KK}^2) \right) \, ,\\
 M^2_{H^\pm}&=&M_H^2\ \left(1 + {\cal O}(M_Z^2/M_{KK}^2)\right)\, ,
 \eea
 where $b_{d_1}$ is an ${\cal O}(1)$ constant depending on the boundary
 conditions on $H(y)$, and in particular on the value of the Higgs brane
 kinetic coefficient $d_1$.

 \section{Minimal Supersymmetric Higgs sector}

 The MSSM is described by the
 superpotential, written in terms of  superfields $\hat{Q}\, ,
 \hat{H_{u}} \, ,  \hat{H_{d}}\, ,  \hat{u^{c}},
 \hat{d^{c}}\hat{L_{i}}$ and $\hat{e^{c}}$: 
\bea
\hspace{-0.7cm}W=Y_{u}^{ij} \hat{Q_{i}} \hat{H_{u}} \hat{u^{c}_{j}} -Y_{d}^{ij}
\hat{Q_{i}} \hat{H_{d}} \hat{d^{c}_{j}} -Y_{e}^{ij} \hat{L_{i}}
\hat{H_{d}} \hat{e^{c}_{j}} + \mu \hat{H_u} \hat{H_d}  \, ,
\eea
where $Y_f^{ij}$ are Yukawa couplings. Two Higgs doublets are
required, one to give masses to up-type quarks, and the other to down-type
quarks and leptons, denoted by $H_u$ and $H_d$ respectively.
 We expand these two doublet complex scalar fields around
their VEVs into real and imaginary parts as
\begin{eqnarray}
\hspace{-.5cm}H_u={ H_u^+ \choose v_u+H^0_u +iA_1 } \, , \ \ 
\hspace{-.1cm}H_d={  v_d+H_d^0+iA_2 \choose H_d^- }\, .
\end{eqnarray}
From the real part of the expansion around $v_u$ and $v_d$, two physical CP-even Higgs states, $h$ and $H$, emerge with the mixing angle $\alpha$ encoding the amount of
$H_u$ and $H_d$ contained within each:
 \begin{equation}
{H \choose h}= \left( \begin{array}{cc}
\cos \alpha & \sin \alpha \\
-\sin \alpha & \cos \alpha   \end{array} \right)
{ H_d^0 \choose H^0_u }\, .
\end{equation}
In a similar way, a CP-odd physical Higgs, $A$, and two charged physical
Higgs bosons,  $H^\pm$, emerge in the spectrum along with their respective
Goldstone bosons. This time, the angle $\beta$, defined as $\tan\beta=v_u/v_d$,
expresses the amount of mixing between the superpotential degrees of freedom and the mass eigenstates:
 \begin{equation}
{G^0 \choose A^0}= \left( \begin{array}{cc}
\cos \beta & \sin \beta \\
-\sin \beta & \cos \beta   \end{array} \right)
{ A_1^0 \choose A^0_2 } \, ,
\end{equation}
and
 \begin{equation}
{G^\pm \choose H^\pm}= \left( \begin{array}{cc}
\cos \beta & \sin \beta \\
-\sin \beta & \cos \beta   \end{array} \right)
{ H_d^\pm \choose H^\pm_u }\, .
\end{equation}
The angles $\alpha$ and $\beta$ are not
independent but are related to one another and can be expressed in terms of the masses of
the physical scalars.
In the MSSM, the Higgs sector can be described by two main input parameters,
commonly chosen to be $M_A$, the pseudoscalar mass, and $\tan \beta$.
The SM-like Higgs mass is predicted to be $M_h=M_Z \cos 2 \beta$ at tree-level,
a relationship that is broken radiatively by loops involving supersymmetric (SUSY) 
parameters, such as stop mixing, and parameters from electroweakino
sector,  all sensitively dependent on the SUSY scale. 

If the SUSY particles are heavy enough to be  out of the
reach of the LHC, the Higgs sector becomes a special case of the Type II 2HDM.
The hMSSM \cite{Djouadi:2013uqa,Djouadi:2013vqa,Djouadi:2015jea}
follows this approach, which has the benefit of removing the explicit
dependence on the SUSY breaking sector (whose main effect is encoded in
the generation of a $125$ GeV Higgs mass).
In this way,
the general effects of SUSY threshold corrections can be accounted for, including
recent MSSM benchmark scenarios that have been proposed in the limit
of heavy SUSY particles (see for example \cite{Bagnaschi:2021jaj}).

In this work we are mostly interested in values of
the new  scalar masses, such that they could be accessible to the LHC, however, indistinguishable from the KK excitations of an extra-dimensional scenario. 
This scenario, with large scalar masses and yet larger SUSY scale, is equivalent to an hMSSM-like approach in the so-called decoupling 
limit ($M_A\gg M_Z$). In that limit, the angles $\alpha$ and $\beta$
are such that \cite{Gunion:2002zf,Djouadi:2013uqa,Djouadi:2013vqa,Djouadi:2015jea} 
\bea
\sin({\beta-\alpha})&=& \ \ 1\ \ \ \ \ \ \  +\   {\cal O}(M_Z^4/M_{A}^4)\, , \\
\cos({\beta-\alpha})&=&\!\!\!  \frac{M^2_Z}{M_A^2}  \frac{\sin (4\beta)}{2}\ \  +\ {\cal O}(M_Z^4/M_{A}^4) \, .
\eea
These relationships are important as Higgs couplings with gauge bosons are
proportional to either $\cos({\beta-\alpha})$ or $\sin({\beta-\alpha})$.
Moreover, in the same limit, the three heavy Higgs bosons have very similar masses, i.e.
\bea
M_A^2 &\equiv&{\cal O}(600-1500\ {\rm GeV})\, ,\\
M_H^2 &=& 
M_A^2\  \left(1 + {\cal O}(M_Z^2/M_{A}^2) \right) \, ,\\ 
M_{H^\pm}^2&=& M_A^2 \left(1+ {\cal O}(M_Z^2/M_A^2) \right)\, . 
\eea
where we consider the pseudoscalar mass $M_A$ to be clearly heavier than the
electroweak scale but not too heavy so that it may still be accessible
at the LHC.

We can see that the Higgs spectrum of the MSSM in the decoupling limit
is actually remarkably similar to the Higgs spectrum of the warped
scenario discussed earlier.

\section{Gauge Couplings - Warped vs MSSM}

We start by comparing the couplings of the different Higgs fields with the SM
gauge bosons, $W$ and $Z$. Both the warped model and the
MSSM contain the usual SM gauge groups $U(1)\times SU(2)$ with gauge
couplings $g_1$ and $g_2$ and with mixing angle $\theta_W$ given by
$\tan{\theta_W}=g_1/g_2$. We define for both models $g_V=(g_Z,g_W)$
with $g_Z=g_2/c_W$ and $g_W=g_2$.
Note that, as explained earlier, in this paper we only consider the
decoupling limit of MSSM ($M_A^2\gg M_Z^2$).
In  Table \ref{tab:cpevenhiggs} we list the couplings of the CP-even Higgs to the gauge bosons, to lowest order approximation, in the two models under investigation.

\begin{table}[h]
  \begin{tabular}{c|c||c|}
    \cline{2-3    }
\multicolumn{1}{c|}{}                   &Warped model\  \ \ & MSSM (decoupling) \ \ \\  
\cline{2-3}
\multicolumn{1}{c|}{}               &   $V_\mu V_\mu$  &   $V_\mu V_\mu$  \\ \hline\hline 
\multicolumn{1}{|c|}{$h$}  & $g_V M_V  $ & $g_VM_V$   \\ \hline
\multicolumn{1}{|c|}{$H$}  &$\varepsilon\ \  g_VM_V\   ({\cal
  I}_V +{\cal I}_H)\ $
& $\varepsilon \ \  g_VM_V \  \sin (4\beta)/2 $    \\ \hline \hline
\multicolumn{1}{|c|}{ $ {\cal O}(\varepsilon)$} &
$\varepsilon= M_Z^2/M_{KK}^2$          & $\varepsilon =M_Z^2/M_{A}^2$   \\ \hline
\end{tabular}
\caption{Trilinear couplings of the CP-even Higgs to gauge bosons to lowest order
  approximation within the MSSM in the decoupling
  limit and minimal warped models containing a single bulk Higgs
  doublet. All order one terms have corrections of 
   ${\cal O}(\varepsilon) $.}
\label{tab:cpevenhiggs}
\end{table}

We see that in both scenarios the lightest CP-even Higgs state has SM-like
couplings with the SM gauge bosons to lowest order.
Also, the couplings of the second CP-even state with
gauge bosons are severely suppressed by
$$
\varepsilon \equiv \varepsilon_{\rm{(Warped~model)}} = M_Z^2/M_{KK}^2  \approx \varepsilon_{\rm{(MSSM)}} = M_Z^2/M_A^2\, , 
$$
where the approximation is only valid in the decoupling limit.\\

In the MSSM case, the coefficients of the suppressed couplings are
well known and, within the decoupling limit, are proportional to $\sin(4\beta)$ \cite{Gunion:2002zf}.
In the warped scenario, due to the orthogonality of eigenstates, some of couplings vanish to the lowest order,  
and the next order corrections
are obtained from the overlap integrals 
\bea
{\cal I}_V &=& \int_0^{y_1}dy e^{-2\sigma} v_5(y) H_1(y)\ \delta
f_V(y),\\
{\cal I}_H &=& \int_0^{y_1}dy e^{-2\sigma} \delta h(y) H_1(y) \, ,
\eea
where $H_1(y)$ is the wave function of the second CP-even Higgs KK
mode, H, along the fifth dimension (the first being the SM Higgs, h).
The terms $\delta f_v(y)$ and
$\delta h(y)$ represent the corrections to the lowest order wave
functions for $h(y)$, $Z(y)$, and $W(y)$ defined as
\bea
h(y) &=& v_5(y) +  \delta h(y)\, ,\\
f_V(y) &=& \frac{1}{\sqrt{y_1+r_1}} +  \delta f_v(y)\, ,
\eea
where  $r_1$ is the coefficient of a possible brane localized gauge kinetic
term (see Eq.~(\ref{eq:5Daction})).

\begin{table}[h]
  \begin{tabular}{cccc|}
\cline{2-4}\cline{2-4}
         & \multicolumn{3}{|c|}{Warped model\ \ \ }\\ \cline{2-4} \cline{2-4} 
         & \multicolumn{1}{|c|}{\ \ \ $V_\mu h$  \ \ \ \ }&\multicolumn{1}{c|}{\ \ \ $V_\mu  H$\ \ \ \ }&\multicolumn{1}{c|}{\ \ \ $V_\mu H^{\pm}$\ \ \ } \\ \hline\hline
\multicolumn{1}{|c|}{$A$}  &  \multicolumn{1}{c|}{$\varepsilon \ \   \frac{g_Z}{2} \ \   ({\cal I}_V +{\cal I}_A)$   }&\multicolumn{1}{c|}{$\frac{g_Z}{2} $\ }   &
\multicolumn{1}{c|}{$\frac{g_W}{2} $\ }\\ \hline
\multicolumn{1}{|c|}{$H^{\pm}$}  &  \multicolumn{1}{c|}{$\varepsilon  \ \  \frac{g_Z}{2} \   ({\cal I}_V+{\cal I}_{H^\pm} )$ }
&\multicolumn{1}{c|}{$\frac{g_W}{2}  $}   &\multicolumn{1}{c|}{$\frac{g_Z}{2} \cos{2\theta_W}$} \\ \hline\hline 
\multicolumn{1}{|c|}{${\cal O}(\varepsilon)$  }  & \multicolumn{3}{c|}{$\varepsilon=M_Z^2/M_{KK}^2$   }    \\ \hline
      &\multicolumn{3}{c}{}\\ \cline{2-4} \cline{2-4} 
%
         &\multicolumn{3}{|c|}{MSSM (decoupling) \ \ \ }\\ \cline{2-4} \cline{2-4} 
         & \multicolumn{1}{|c|}{\ \ \ $V_\mu h$  \ \ \ \ }&\multicolumn{1}{c|}{\ \ \ $V_\mu
  H$\ \ \ \ }&\multicolumn{1}{c|}{\ \ \ $V_\mu H^{\pm}$\ \ \ } \\ \hline\hline
\multicolumn{1}{|c|}{$A$}
& \multicolumn{1}{c|}{$\varepsilon \ \  \frac{g_Z}{2}  \ \   \sin (4\beta)/2 $ }
& \multicolumn{1}{c|}{$\frac{g_Z}{2}  $\ }
& $\frac{g_W}{2} $\ (exact)    \\ \hline
\multicolumn{1}{|c|}{$H^{\pm}$}
&\multicolumn{1}{c|}{$ \varepsilon \ \   \frac{g_W}{2}   \   \sin (4\beta)/2 $ }
&\multicolumn{1}{c|}{$\frac{g_W}{2}$}
& $\frac{g_Z}{2}\cos{2\theta_W}$\  \\ \hline\hline 
\multicolumn{1}{|c|}{${\cal O}(\varepsilon)$  }       & \multicolumn{3}{c|}{$\varepsilon=M_Z^2/M_{A}^2$ }                                   \\ \hline
\end{tabular}
\caption{Trilinear couplings of gauge fields with CP-odd and charged Higgs bosons, to lowest order
  approximation within the MSSM in the decoupling
  limit, and minimal warped models containing a single bulk Higgs
  doublet. Except for the CP-odd coupling to charged Higgs bosons in the MSSM case, all order one terms have corrections of 
   ${\cal O}(\varepsilon) $.}
\label{tab:cpodhiggs}
\end{table}

Similarly, for the case of the CP-odd Higgs, $A$, and the charged  Higgs bosons, $H^{\pm}$,
the trilinear couplings with the gauge fields have the same behaviour in both BSM
scenarios, with the same type of suppression. The couplings are listed in Table \ref{tab:cpodhiggs}.
The suppressed couplings of the MSSM in the decoupling limit are again 
 proportional to $\sin(4\beta)$, while in the warped
scenario two new overlap integrals must be defined, which take into
account the EWSB corrections to the wave functions of the CP-odd and
the charged Higgs bosons. The overlap integrals are
\bea
{\cal I}_A &=& \int_0^{y_1}dy e^{-2\sigma} v_5(y)\ \delta A(y) \, ,\\
{\cal I}_{H}^\pm &=& \int_0^{y_1}dy e^{-2\sigma} v_5(y) \delta H^{\pm}(y)\, ,
\eea
containing the wave functions of the heavy Higgs bosons along the bulk of the extra dimension
\bea
A(y) &=& H_1(y) + \delta A(y)\, ,\\
H^{\pm}(y) &=& H_1(y)+  \delta H^{\pm}(y)\, ,
\eea
where it is explicit that in the absence of EWSB, the wave functions
of $H$,  $A$, and $H^\pm$ are the same.

\section{Yukawa couplings - Warped vs MSSM}
For the warped scenario,
 we will consider {\it bulk} Yukawa coupling operators as well as {\it brane}
localized operators. Because localization of fermion fields is responsible for
their hierarchical masses and mixings \cite{Agashe:2004cp},  the usual flavour paradigm of 5D
warped scenario will be unchanged by the presence of both bulk and brane
Yukawa operators.
 However, the couplings between SM fermions and
heavy Higgs bosons can be very sensitive to whether their corresponding Yukawa terms are brane localized or propagating in the bulk.

We consider the following 5D quark Yukawa Lagrangian density:
\bea
&&\hspace{-1cm} {\cal L}_{Y}= \sqrt{g} \left( \frac{Y_u^{bulk}}{\sqrt{k}}  HQU +
  \frac{Y_d^{bulk}}{\sqrt{k}} HQD \right)\non\\ 
&&\hspace{-1.5cm}+  \sqrt{g}\, \delta(y-y_1) \left( \frac{Y_u^{brane}}{\sqrt{k}}  HQU +
\frac{Y_d^{brane}}{\sqrt{k}} HQD\right) + h.c. \!,
\label{5DYukawas}
\eea
where $Q$, $U$, and $D$ denote 5D quark doublet and singlets of
$SU(2)$ respectively. We have also introduced the dimensionless bulk and
brane Yukawa couplings $Y^{bulk}_i$ and $Y^{brane}_i$, and the 5D
Higgs field as $H$\footnote{In the case of CP-odd and charged Higgs bosons, there
  will be a subdominant gauge coupling contribution coming from the 5D covariant
  derivative, since the physical CP-odd and charged scalars are 
  admixtures of bulk Higgs and gauge fields. The contribution will
  appear after the diagonalization of the (infinite) fermion mass
  matrices and will be suppressed by ${\cal O}(v^2/M^2_Q)$, where $M_Q$
  is the mass of the first fermion KK mode. We will neglect these
couplings in the remainder of this work.}.

\begin{table}[t]
\begin{tabular}{|c|c|c|}
\cline{1-3}
     & Type II Warped    & MSSM   \\ 
     & model & (decoupling) \\    \hline\hline 
 $ y_{hqu}$& $(m_u/v)$ &  $(m_u/v)$   \\ \cline{1-3}
 $y_{hqd}$  & $(m_d/v)$ &  $(m_d/v)$  \\ \hline\hline
$y_{Hqu}, y_{Aqu}, y_{{H^\pm}qu}$  &$(m_u/v) \frac{H(y_1)}{h(y_1)}$ &  $(m_u/v) \cot{\beta}$    \\ \cline{1-3} 
$y_{Hqd}, y_{Aqd}, y_{{H^\pm}qd}$ & $(m_d/ v) { B_d} $ &   $(m_d/v) \tan{\beta}$   \\ \hline\hline
 $\times \Big(1+{\cal O}(\varepsilon)\Big)$  &\!\!\! $\varepsilon=
\frac{v^2}{M_{KK}^2}\, {\rm and/or}~\delta Y^{i}$ \!\!\!          & $\varepsilon=M_Z^2/M_{A}^2$   \\ \cline{1-3}
\end{tabular}
\caption{Lowest order approximation of the Yukawa couplings between $up$ and $down$ SM quarks and the SM-like
  Higgs, $h$, the heavy CP-even Higgs, $H$, the  CP-odd Higgs, $A$, and
  the charged Higgs, $H^{\pm}$, within the MSSM in the decoupling
  limit and minimal warped model described here. For
  simplicity, we have only considered one generation, and thus suppressed
  any mixing angle dependence in the heavy Higgs couplings.}
\label{tab:yukawa}
\end{table}

The Yukawa
couplings between the SM-like Higgs and up-type and down-type quarks will be
obtained to lowest order from the overlap integrals between the light Higgs boson
and the SM fermion wave functions along the bulk of the extra-dimension (note that higher order corrections could
lead to visible effects in flavour physics  \cite{Azatov:2009na}).
With this structure, the SM quark Yukawa couplings become
 \bea
  y^{SM}_{i} &=& \frac{Y_i^{bulk}}{\sqrt{2k}} \int_0^{y_1} dy e^{-4 \sigma}F_q(y)
  F_i(y) h(y)  \non\\
  &&+ \frac{Y_i^{brane}}{\sqrt{2k}} e^{-4 \sigma(y_1)}F_q(y_1)
  F_i(y_1) h(y_1)  \, ,
\eea
where $i=u,d$ and with the fermion wave functions given by
$F_q(y)= N_q\ e^{(2-c_q) 
  \sigma(y)}$ and $F_i=N_i\ e^{(2+c_i) \sigma(y)}$, such that $N_q$
and $N_i$ are the appropriate fermion canonical normalization factors.
This can be rewritten as
\bea
  \hspace{-0.7cm}y^{SM}_{i} = \frac{N_qN_i h(y_1)}
{\sqrt{2k}}e^{(\Delta c) \sigma(y_1)}
\left( Y_i^{bulk} F_h(\Delta c) +Y_i^{brane}\right) \, ,
  \eea
  where we have defined
  \bea
  F_h(\Delta c)\equiv \int dy e^{(\Delta c) (\sigma(y)-\sigma(y_1))}
  h(y)/h(y_1) \, ,
  \eea and $\Delta c = c_i-c_q$.
These couplings will be hierarchical because of their sensitivity on
$\Delta c$, and thus will lead to hierarchical quark mass matrices
given by $m_i = y^{SM}_{i}\ v $.

We can obtain the heavy Higgs bosons and SM fermions Yukawa couplings in the same fashion, for example in
the case of the Heavy  CP-even Higgs Yukawa couplings we have  
  \bea
 \hspace{-0.7cm}  y_{Hqi} =
    \frac{N_qN_i H(y_1)}
{\sqrt{2k}}e^{(\Delta c) \sigma(y_1)}
\left( Y_i^{bulk} F_H(\Delta c) +Y_i^{brane}\right) \, ,
\eea
where  
\bea
F_H(\Delta c)= \int dy e^{\Delta c
  (\sigma(y)-\sigma(y_1))}H(y)/H(y_1)\, .
\eea
 The Yukawa couplings of the fermions with the Higgs bosons in the two scenarios considered here are listed in Table \ref{tab:yukawa}.

\begin{figure}[t]
\center
\includegraphics[height=6.7cm,width=8.3cm]{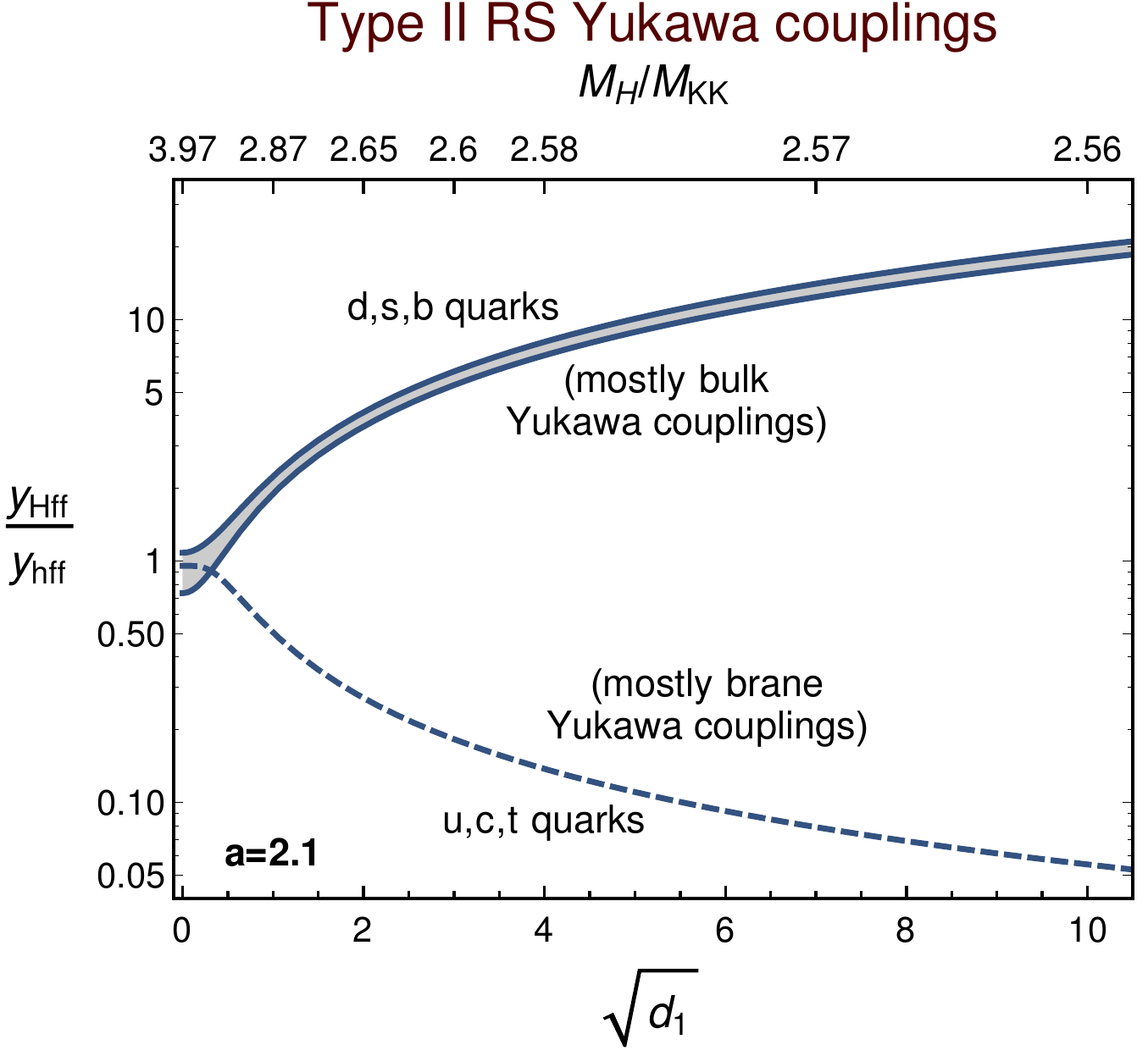}

\vspace{.3cm}
  
  \includegraphics[height=6.5cm,width=8.5cm]{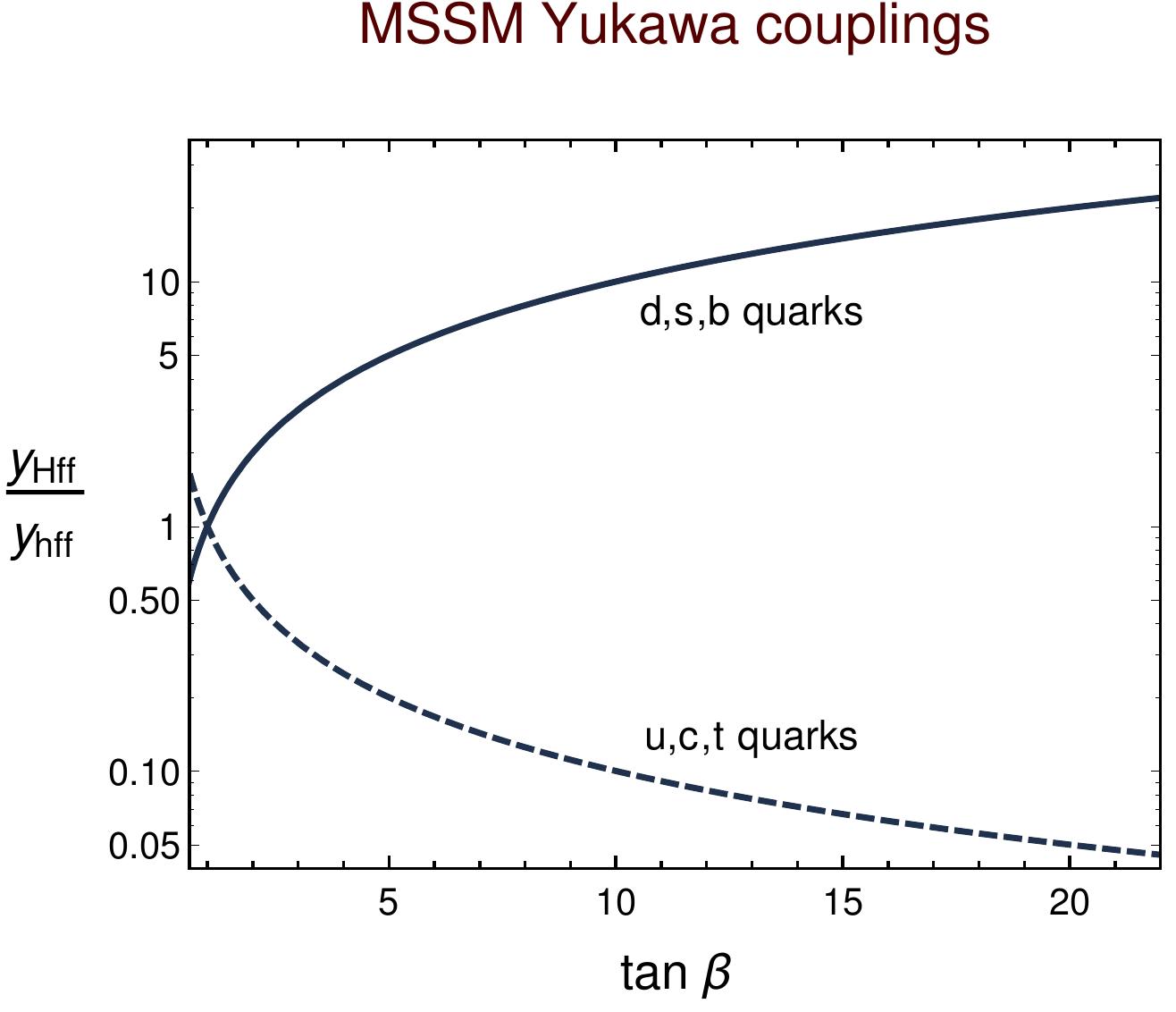} 
\vspace{-.2cm}
\caption{Ratio of Yukawa couplings $\hat{y}_{Hff}/y_{hff}$ in the warped scenario with 
RS metric
(top panel) and  in the MSSM (bottom panel).  
The parameters
responsible for splitting the couplings are $\tan\beta$ in the MSSM
and $d_1$ (the coefficient of the IR-brane localized Higgs kinetic
term) in the warped scenario.}
\label{fig:yvsrh}
\vspace{-.2cm}
\end{figure}

In principle 
there is no reason for the bulk and brane Yukawa coefficients to be of different orders,
however, we envision here the case where a relative hierarchy between the two
contributions exists.

First we assume that within the up-sector we have $Y_u^{bulk}
\ll Y_u^{brane}$, so that $\delta Y^u_{h,H} =
\frac{F_{h,H}Y_u^{bulk}}{Y_u^{brane}} \ll 1$. In that limit we obtain 
\bea
\frac{y_{Hqu}}{y_{hqu}} = \frac{H(y_1)}{h(y_1)} \left(1 + {\cal O} (\delta
Y^u) \right).
\eea
We can further consider the simple limit in which the metric background is
the RS metric with the warp factor given by $\sigma(y) = k y$, where $k
\approx M_{Pl}$,
and the bulk Higgs potential is quadratic in $H$, i.e. we have
$V(H)=a (a-4) k^2 H^2$. Here, $a$ is a parameter bound by $a>2$ to
ensure that the Higgs VEV, $v(y) \propto e^{ a k y}$,  is sufficiently localized towards the IR brane
in order to address the hierarchy problem. In the limit of large Higgs
brane kinetic term coefficient, $d_1$, we obtain the asymptotic behaviour:
\bea
 \frac{H(y_1)}{h(y_1)} \sim \frac{M_{KK}}{M_H} \sqrt{\frac{2}{d_1}}
 \ \ \  \ ({\rm large\  d_1,\ RS\ metric}) \, .
 \eea
Note this relationship is flavour independent as it does not depend on the
structure of the 5D bulk mass parameters.

We next assume that for the down-sector quarks, 
the Yukawa couplings hierarchy is inverted so that $Y_d^{bulk}  \gg
Y_d^{brane}$. With  the new small parameters $\displaystyle \delta Y^d_{h,H}
=\frac{Y_d^{brane}}{F_{h,H}Y_d^{bulk}} \ll 1$ we can again expand
the ratio of Yukawa couplings as
\bea
\frac{y_{Hqd}}{y_{hqd}} =B_d(\Delta c)
\left(1 + {\cal O} (\delta
Y^d) \right) \, ,   
\eea
where we have defined
\bea
B_d(\Delta c)= \frac{\int dy
  e^{(\Delta c) \sigma(y)}H(y) 
}{\int dy e^{(\Delta c)  \sigma(y)} h(y)}\, ,
\eea
and again $\Delta c=c_u-c_q$.
When the background metric is RS and the bulk Higgs
potential is quadratic, one could obtain a closed form solution for the integral
$B_d$ in terms of hypergeometric functions. This solution has been used
in Figures 1 and 2.
In general, and for large values of $d_1$ the Yukawa coupling scales as
\bea
\frac{y_{Hqd}}{y_{hqd}}  \sim \sqrt{d_1}\times {\cal O}(1) \ \ \ ({\rm large}\ d_1),
\eea
which shows the inverse dependence on $d_1$ compared to the
mostly-brane Yukawa coupling (that we propose for the up-sector). We show the comparison of the warped scenario Yukawa couplings with those of the MSSM in the decoupling limit in Fig. \ref{fig:yvsrh}, top and bottom panels, respectively.

\begin{figure}[h]
\center
\begin{center}
  \includegraphics[height=7cm,width=8.8cm]{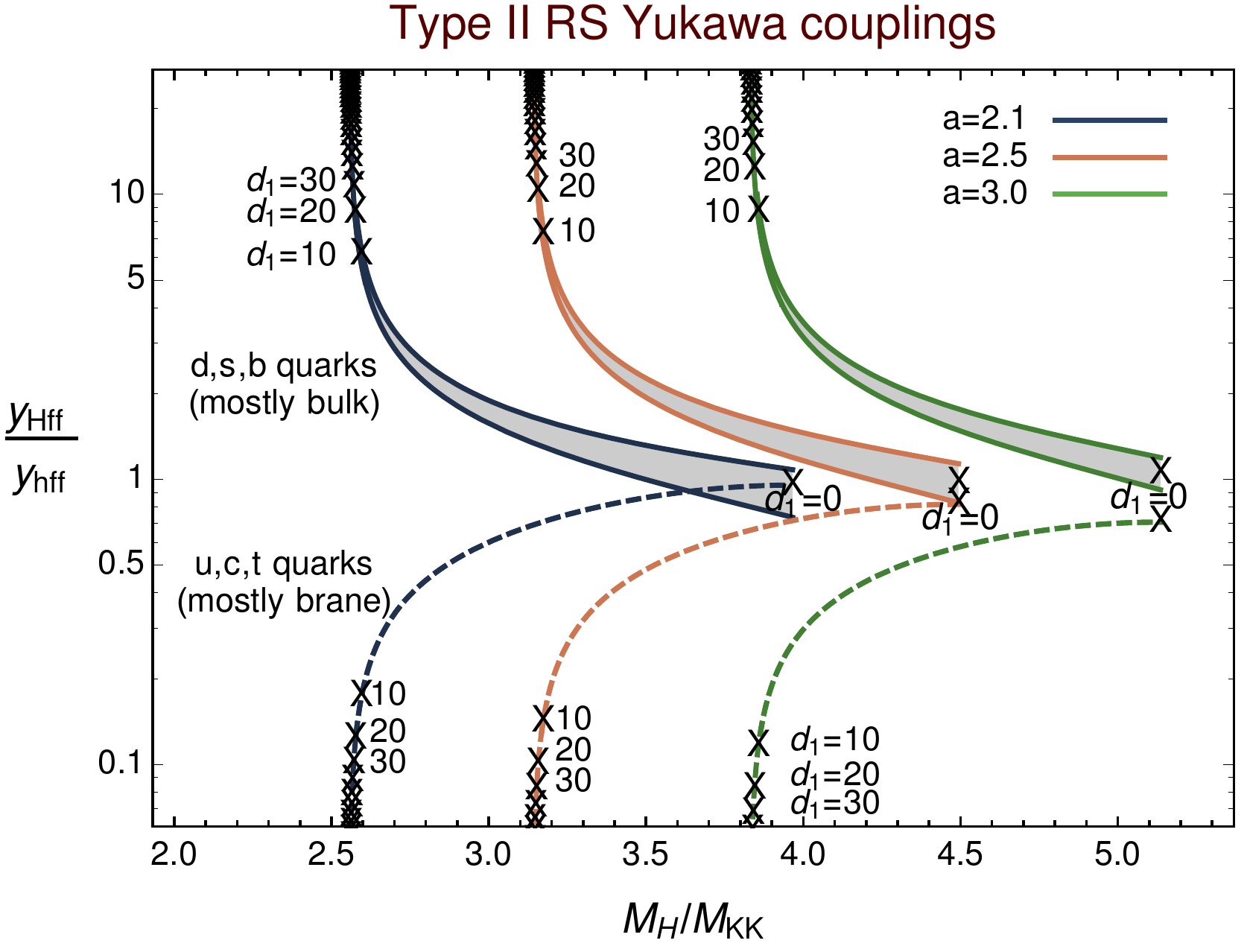}\ \ 
\end{center}
\vspace{-.2cm}
\caption{Ratio of Yukawa couplings, $y_{Hff}/y_{hff}$, in the warped
scenario  
with the RS metric.  
We scan with respect to the heavy Higgs
mass $M_H$ (relative to the KK scale $M_{KK}$), and show some 
of the corresponding values of the Higgs brane kinetic term
coefficient, $d_1$. Three different values of the bulk Higgs mass parameter, $a$, given in the legend at the top right corner, are
considered.}
\label{fig:hatyvsrH}
\vspace{-.2cm}
\end{figure}

It can be shown that to lowest order in $v^2/M_{KK}^2$, the couplings of
the CP-even $H$, the CP-odd $A$, and the charged Higgs $H^{\pm}$ bosons are
actually the same (all of these KK excitations originate from the same 5D
Higgs doublet and thus their couplings come from the same 5D Yukawa
operators). We thus write
\bea
&&y_{Hqu}\simeq y_{Aqu} \simeq y_{H^\pm qu}\, ,\\
&&y_{Hqd}\simeq y_{Aqd}\simeq   y_{H^\pm qd}\, ,
\eea
where $y_{H^\pm qu}$ is the coefficient appearing in the coupling between left-handed
down quark and right-handed up quark, and  $y_{H^\pm qd}$ corresponds to
left-handed up quark and right-handed down quark. Also note that when
the full flavour structure is considered, the heavy Higgs couplings of
the warped scenario can involve off-diagonal entries due to a misalignment
  between the quark mass matrix and the coupling matrix between heavy
  Higgs bosons and quarks. This effect does not appear within the MSSM as
  the couplings are simultaneously diagonalized with the quark
  matrices, and only the charged Higgs couplings involve the CKM
  entries in a very straightforward way.
The collider phenomenology will involve mainly the heavy flavours (top
and bottom quarks) so we focus here on these couplings. The ratio of the Yukawa
couplings of the top and bottom quarks with the heavy Higgs bosons versus the SM Higgs
bosons is plotted in Fig. \ref{fig:hatyvsrH}.

The first limit, in which the  5D Yukawa couplings are
dominated by the brane Yukawa operator, is interesting because it
shows a linear dependence on the value of the wave function of the
heavy Higgs at the TeV boundary. This value  can
be controlled with the Higgs brane kinetic term, which suppresses it
for large brane kinetic coefficient, $d_1$. This way one can
suppress the coupling of the heavy Higgs with up-type fermions, mimicking the
behaviour in the MSSM scenario for large $\tan \beta$.
For down-type fermions, the Yukawa couplings of the heavy Higgs bosons have a
more obscure dependence, but they show a clear enhancement for large
brane kinetic coefficient $d_1$.

Of course, in general, the warped scenario can have any Yukawa
structure. We call the flavour structure in which the up(down)-type fermions have
dominant brane(bulk) Yukawa couplings, a Type II Yukawa coupling setup, as it is
reminiscent of the Type II  2HDM and the
MSSM. Note that other flavour structures are possible in the warped
  setup and these can easily resemble other types of
  Yukawa couplings of specific 2HDM's, such as Type I or flipped Yukawa structures.

\section{Discussion and outlook}

We have shown that the gauge couplings of heavy exotic Higgs fields
in minimal implementations of both supersymmetry and warped extra
dimensions are very similar when the masses of the new Higgs bosons are
significantly heavier than the electroweak scale and that this is a general result. 

We have also shown that the Yukawa couplings of the exotic Higgs bosons can also
be very similar in a specific parameter region of the warped extra
dimensional model.

The production cross section of neutral Higgs bosons through gluon fusion
is governed by their Yukawa couplings. Thus both
minimal BSM scenarios analyzed here could yield very similar production
rates. The production of charged Higgs bosons will go directly through Yukawa
couplings and again we could have very similar production rates in both scenarios.

Once these exotic Higgs bosons are produced they will decay, via gauge
 (or Yukawa) interactions. Since gauge couplings are
similar in both models, it is then quite possible to have a regime in which the 
whole phenomenology of exotic Higgs bosons is indistinguishable in the
two different types of scenarios.

This however does not mean that they have to be always indistinguishable, since
the parameter space of warped extra dimensions in the Yukawa sector is
much larger. When the Yukawa coupling structure is not governed by 
hierarchies within brane and bulk Yukawa operators, one would expect both BSM 
models to be similar {\it only} when the top quark coupling to the heavy Higgs bosons
dominate over the bottom quark couplings, {\it i.e.} in the case of $\tan
\beta$ of order 1 in the MSSM. 

We have here focused on the simplest metric background for
the warped extra-dimensional model. However minimal implementations of
the SM within the RS background have strong constraints from precision
electroweak tests and flavour phenomenology, and 
it is known  that many of these bounds can be relaxed in the presence
of slightly modified metric backgrounds \cite{Cabrer:2011fb}, without
changing the minimal structure of the Higgs sector. Moreover it was
also shown that within these modified scenarios, the heavy Higgs modes
can easily have masses well below the masses of the rest of KK
excitations of the model \cite{Frank:2016vtv}. 

Thus one can perform more realistic phenomenological comparisons of the two models within these new metric scenarios.
The disadvantage is that, in that case,  analytical expressions for
the couplings cannot be obtained in general and one has to rely on numerical computations. The study
performed here with the RS metric thus is a preliminary
first order study, important in its transparency, and useful for checking the numerical results emerging 
from a more realistic scenario, which will be the subject of further
studies.

\section{Aknowlegements}

The work of M. F.  has been partly supported by NSERC
through the grant number SAP105354.
M. T. would like to thank FRQNT for financial support under              
grant number PRC-290000. This paper reflects solely the authors' personal opinions and does not represent the opinions of the authors' employers, present and past, in any way.

\bibliographystyle{elsarticle-num-names} 
\bibliography{warped}

\end{document}